\documentclass[transmag]{IEEEtran}
\usepackage{latexsym}
\usepackage{graphicx}
\usepackage{amsfonts,amssymb,amsmath}
\usepackage{hyperref}
\usepackage{cite}
\usepackage{gensymb}
\usepackage{xcolor}
\usepackage{ulem}
\newcommand{\abs}[1]{\left\lvert #1 \right\lvert }

\def\BibTeX{{\rm B\kern-.05em{\sc i\kern-.025em b}\kern-.08em T\kern-.1667em\lower.7ex\hbox{E}\kern-.125emX}}
\begin{document}

\title{A Reconfigurable Intelligent Surface at mmWave based on a binary phase tunable metasurface }

\author{Jean-Baptiste Gros, Vladislav Popov, Mikhail A. Odit, Vladimir Lenets and  Geoffroy Lerosey,
\thanks{Greenerwave  acknowledge funding from the French ``Minist\`{e}re des Arm\'{e}es, Direction G\'{e}n\'{e}rale de l'Armement'' and  ``Agence de l’Innovation de d\'{e}fense''  through the RAPID  ``3SFA'' project. and the European Commission through the H2020 project through the RISE-6G, HEXA-X (Grant Agreement no. 101015956) }
\thanks{J.-B. Gros was with ESPCI Paris, PSL University, CNRS, Institut Langevin 75005 Paris, France. He is now with Greenerwave, 6 rue Jean Calvin, 75005 Paris, France (e-mail: jean-baptiste.gros@espci.fr)  }
\thanks{V. Popov is with Greenerwave, 6 rue Jean Calvin, 75005 Paris, France (e-mail: vladislav.popov@greenerwave.com)}
\thanks{M. A. Odit, is with Greenerwave, 6 rue Jean Calvin, 75005 Paris, France (e-mail:mikhail.odit@greenerwave.com)} 
\thanks{V. Lenets is with ESPCI Paris, PSL University, CNRS, Institut Langevin 75005 Paris, France (e-mail:vladimir.lenets@espci.psl.fr)} 
\thanks{G. Lerosey is with Greenerwave, 6 rue Jean Calvin, 75005 Paris, France (e-mail: geoffroy.lerosey@greenerwave.com)}
}

\IEEEtitleabstractindextext{\begin{abstract} 
Originally introduced in the early 2010's, the idea of smart environments through reconfigurable intelligent surfaces (RIS) controlling the reflections of the electromagnetic waves has attracted much attention in recent years  in preparation for the future 6G. Since reconfigurable intelligent surfaces are not based on increasing the number of sources, they could indeed pave the way to greener and potentially limitless wireless communications. In this paper, we design, model and demonstrate experimentally a millimeter wave reconfigurable intelligent surface based on an electronically tunable metasurface with binary phase modulation. We first study numerically the unit cell of the metasurface, based on a PIN diode, and obtain a good phase shift and return loss for both polarizations, over a wide frequency range around 28.5~GHz. We then fabricate and characterize the unit cell and verify its properties, before fabricating the whole 10~cm$\times$10~cm reconfigurable intelligent surface. We propose an analytical description of the use that can be done of the binary phase RIS, both in the near field (reflectarray configuration) and in the far field (access point extender). We finally verify experimentally that the designed RIS works as expected, performing laboratory experiments of millimeter wave beamforming both in the near field and far field configuration. Our experimental results demonstrate the high efficiency of our binary phase RIS to control millimeter waves in any kind of scenario and this at the sole cost of the energy dissipated by the PIN diodes used in our design. 
	\end{abstract}

}

\maketitle 
\section{INTRODUCTION}

\IEEEPARstart{W}{ith}  5G being currently under deployment, wireless communication engineers are very close to have harnessed and improved the whole chain of wireless systems. Indeed, during the past generations, modulation and coding techniques have been largely optimized to maximize the data rates, MIMO ideas have been introduced to multiply capacities, and multiple bands have been aggregated to maximize communication bandwidths\cite{Foschini1998,paulraj2003introduction,Tarokh1998,Alamouti1998}. With 5G, even more bricks are being introduced, with both millimeter wave (mmWave) frequencies offering ultrawide bandwidths and Massive MIMO antennas bringing in spatial multiplexing, to offer very high data rates to the end users\cite{Marzetta2013, Larsson2014,Lu2014,Bjornson2015,Bjornson2016}. It is to be noted that these new performances almost always come at the price of increased hardware complexities and higher energy consumption.

Looking at it from this perspective, the horizon may seem somehow technologically cluttered. Yet there is a missing part of paramount importance that has been kept aside: the channel itself. This is why the idea of smart environments, that can adapt themselves for superior wireless communications in real time, has attracted much attention in the context of 6G. Being able to modify the channel at will, with energy-sober technologies, could indeed pave the way to greener and potentially limitless wireless communications. 

The concept of smart environments was first introduced in \cite{Subrt1,Subrt2}, where the authors proposed to optimize the usage of various access points inside a building, by controlling the electromagnetic reflectiveness of its walls, in order to distribute the signal in an optimal way in the building according to the needs.
A bit later, the idea of using electronically reconfigurable surfaces for improved wireless communications was proposed in \cite{Kaina2014,Dupre2015}, where the fields on the surface were locally controlled in order to focus on a given antenna. 
Some follow up works were published over the years, such as \cite{X.Tan,Welkie,Hougne}, but the topic remained of rather little interest to the community, until a series of papers were published in late 2018 and 2019 \cite{Hougne,Liaskos_7,Liaskos_8,Renzo2019,Basar2019}. 
From there, the idea of smart environments using electronically reconfigurable surfaces, coined Reconfigurable Intelligent Surfaces (RIS) emerged as a credible major evolution for 6G and has attracted an enormous and growing interest in the wireless communications community. Associated to RIS there are numerous research topics ranging from energy efficient wireless communication\cite{8741198} to channel modeling \cite{DiRenzo2020,Garcia2020}, RIS based signal modulation and encoding \cite{Basar2020},  MIMO channel estimation and beamforming \cite{Nadeem2020,He2020,Park2020}, telecommunication  performance evaluation \cite{Badiu2020},  mathematical model  and  optimization method for wavefront shaping with RIS   \cite{Wu2020,Zappone2020,Huang2019} and even stochastic analysis approach \cite{DiRenzo2019}. 
Yet of the numerous works proposed, a very large part is concerned with theoretical and mathematical approaches, and very few deal with experimental demonstrations of RIS. 
This lack of practical demonstrations is even worse in the context of millimeter waves, where RIS first use case may well be found for instance as passive access point extenders.

In this paper we propose an implementation of a low complexity RIS at mmWave and show its potential for wireless communications. To do so, we demonstrate an electronically tunable binary phase metasurface, able to control the reflections of electromagnetic waves in the 30GHz range, at a half wavelength pitch and with a $\pi$-phase shift on two polarizations independently. We first start by explaining the design of the unit cell, which is based on a patch reflector capacitively coupled to a parasitic resonator controlled by a PIN diode, as proposed in \cite{Kaina:14}. 
In contrast to other reconfigurable metasurfaces, which are often built from a patch with a PIN diode connected to the RF ground~\cite{5765450,7448838,7902179},
in our design we control the reflection phase by means of a capacitively-coupled to a patch parasitic resonator loaded with a PIN diode.
Although this approach is more challenging at mm-Waves, it provides us more degrees of freedom for optimizing the design to significantly increase the instantaneous bandwidth and to reduce power dissipation in active elements.
We study the properties and performance of the unit cell using commercially available FEM software. Then, we fabricate the metasurface  unit cell, characterize its properties using standard laboratory equipment, and verify that we can obtain a $\pi$-phase shift when changing the state of the PIN diode (forward or biased polarized). We show dissipation levels of the unit cell below 2dB, for both polarizations of the unit cell. We finally provide a description of the final assembled RIS, that contains $20\times20$ unit cells on a $10$~cm~$\times$~$10$~cm surface, with a total of 800 PIN diodes controlled by an FPGA board. In the next part of the manuscript, we derive an analytical model of   binary phase RIS. This model is then used to investigate  two physical scenarios of a wave  impinging the RIS namely (i) a spherical wave and (ii) a plane wave. These scenarios correspond to the  two practical uses of the binary phase RIS  in the near field (reflectarray configuration) and in the far field (access point extender). This analytical investigation underlines the peculiarities of binary phase shifting metasurfaces when dealing either with spherical or  planes waves, and explain their mechanism. Finally, in the last part of the paper, we propose to experimentally explore both physical situations. In a first experiment, we use the  RIS, that we have designed,  in a near field configuration, corresponding to a typical reflectarray antenna. In this context, we demonstrate  experimentally that the RIS can be used to beamform mmWaves very efficiently, achieving directivity close to 30dBi, and reaching angles as high as $60^\circ$ as predicted by the analytical model. The last experiment leverages the RIS in a far field configuration and provides an experimental proof of wireless transmission of mmWaves in a scenario where two horn antennas connected to a VNA don’t have any line of sight. We demonstrate how the RIS can be used to create such a line of sight, hence providing a 25dB gain over the received energy.

\section{ Binary reconfigurable intelligent surface}
\subsection{Unit cell design}
\begin{figure}
\centerline{\includegraphics[width=3.5in]{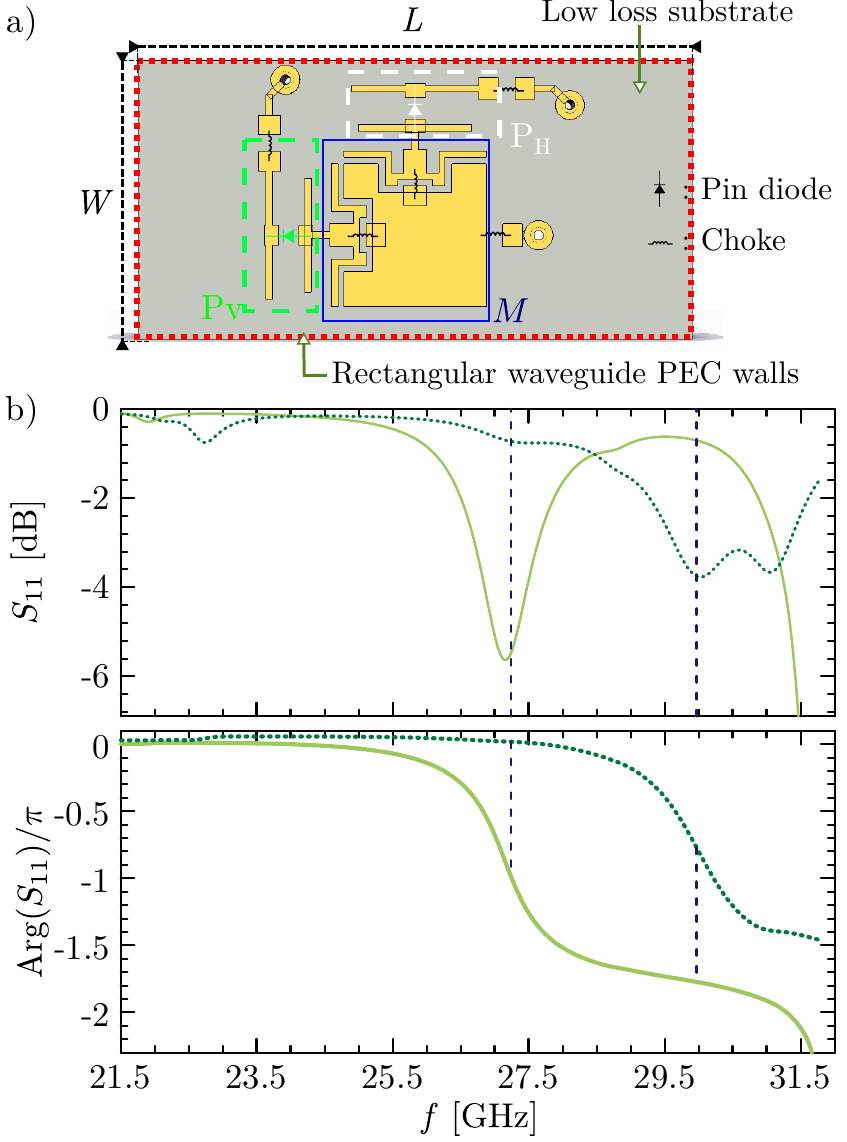}}
\caption{\label{sec1:fig1} a) Unit cell of the designed reflect array. $M$ - main patch, $P_H$ and $P_V$ are horizontal and vertical parasitic resonators, respectively. When studied experimentally the pixel is placed inside the WR-34 hollow waveguide. b) Simulated reflection from the metasurface unit cell. Solid line represents OFF-state and doted line represents ON-state of the pixel. Vertical dashed lines show the operating frequency range of the pixel.}
\end{figure}

The single unit cell (or pixel) of the designed metasurface (\ref{sec1:fig1}) is represented by a conducting square patch (made of copper) placed on a low loss grounded dielectric substrate Meteorwave 8000 ($\varepsilon=$ 3.3 @ 10~GHz, $tan(\delta)=$ 0.0016 @ 10~GHz)\cite{meteorwave}. 
At the resonance frequency of the patch the incident wave strongly interacts with the unit cell leading to enhanced power dissipation and significant phase advance of the reflected wave.
The unit cell is designed to provide the patch resonance $f_0$ to be approximately  around 27.5 GHz. 

In order to provide the proper operation of the binary reflect array, two conditions should be fulfilled: the amplitude of the reflected wave should be maximum and the reflection phase should be able to switch in 2 states with a $\pi$-phase difference. There are many ways to change the phase of the reflected wave and most of them are based on changing the electrical length of the resonator. The simplest way to do it is to shunt the patch resonator onto the ground using an active element i.e. PIN diode. Nevertheless this leads first to a strong dissipation growth due to increased current through the PIN diode.  Moreover this system becomes quite sensitive to the fabrication intolerances of the diodes. 
Due to this reason here we use a different method to control reflection phase. In order to provide a $\pi$-phase shift, the design of the pixel is complemented with a parasitic resonator ($P_H$ and $P_V$ on Fig.\ref{sec1:fig1}a for both corresponding horizontal and vertical polarizations of the E-vector of incident wave). The resonance frequency of the parasitic resonator lies close to the resonance of the patch. That leads to appearance of strong coupling between two resonators and corresponding anti-crossing behavior when coupled resonances repulse and the mutual resonance shifting occurs. While changing the electrical length of the parasitic resonator we change its resonance frequency and can tune 
the mutual coupling between resonant modes of the parasitic resonator and the patch.
This in turn allows to change the resonance frequency of the patch and the phase of the reflected wave at a given frequency. This method was first suggested and verified in \cite{Kaina:14}.
In order to provide control of the electrical length of the parasitic resonator, it is split into two parts connected with a PIN diode (MACOM-000907). To make the  unit cell more compact the parasitic resonator is partially placed inside the preliminary carved patch resonator. The diode state is switched via applied voltage or current, controlled through the biasing network. The network is isolated from the RF-part of the unit cell through the lumped inductive elements (chokes). To increase the structure symmetry, one of the chokes of the parasitic resonator which is closest to patch is placed in the middle of the resonator and is connected to the patch. The patch is then connected with the other pole of the control network through the common patch choke, which is placed on the right side of the patch (see Fig.~\ref{sec1:fig1}.a). Electrical connection to the biasing network between top and bottom layers is provided through the vias. All the control electronics is placed on the back side of the multilayer PCB. When we apply a 5V biasing voltage, the state of the PIN diode is changed from isolating (we call it \textit{OFF-state}) to conducting (\textit{ON-state}). The electrical length of the parasitic resonator changes correspondingly providing shifting of its own  resonance frequency and consequently of the resonance frequency of the main patch.
Fig.~\ref{sec1:fig1}.b) shows simulated reflection coefficient from the infinite periodic structure composed of the described pixels. The simulation was performed with the time domain solver of the CST Studio Suite software. 
The unit-cell is simulated in the rectangular waveguide (WR-34) as illustrated by Fig.~\ref{sec1:fig1} a). A single-mode rectangular waveguide port is set two wavelength (at 28.5~GHz) away from the top of the unit cell. However, the simulation is performed in a wide frequency range and at the lowest frequency the distance to the port is approximately one wavelength. The port is de-embedded having the reference plane at the top of the unit cell in order to subtract the propagation phase.
In the OFF-state the resonance of the patch (27.5~GHz, in Fig.~\ref{sec1:fig1}.b)) lies to the left of the resonance of the parasitic resonator (30~GHz  at Fig.~\ref{sec1:fig1}.b)). In the ON-state the electrical length of the parasitic resonator increases and the parasitic resonance is shifted to the lower frequency (22.5~GHz). The coupled patch resonance is correspondingly pushed to the higher frequency (29.5~GHz). 
The operating frequency range of the unit cells  is defined from the acceptable level of the reflection coefficient amplitude and difference of the reflection coefficient phases between ON- and OFF-states. 
To operate, a RIS is required to modify the phase of the reflected wave in comparison to the incident wave and not to dissipate all the incident power. 
The maximum phase difference possible is 180 degrees, but any value close to it is enough.
For the designed pixel shown in Fig.\ref{sec1:fig1}.a), the operating frequency range lies in the frequency band from 27.5~GHZ to 29.5~GHz. 
In this range, marked with the two vertical dashed lines in Fig.\ref{sec1:fig1}.b), independently of the pixel's state, the phase difference is always larger than 160 degrees.
Meanwhile, the reflection amplitude level never goes under the -5~dB level, meaning that more than 20\% of power is reflected. 
Although these characteristics are not perfect and can be further improved, we demonstrate in what follows that they are acceptable for the target applications.

\subsection{Unit cell measurements}
\begin{figure}
	\centerline{\includegraphics[width=3.5in]{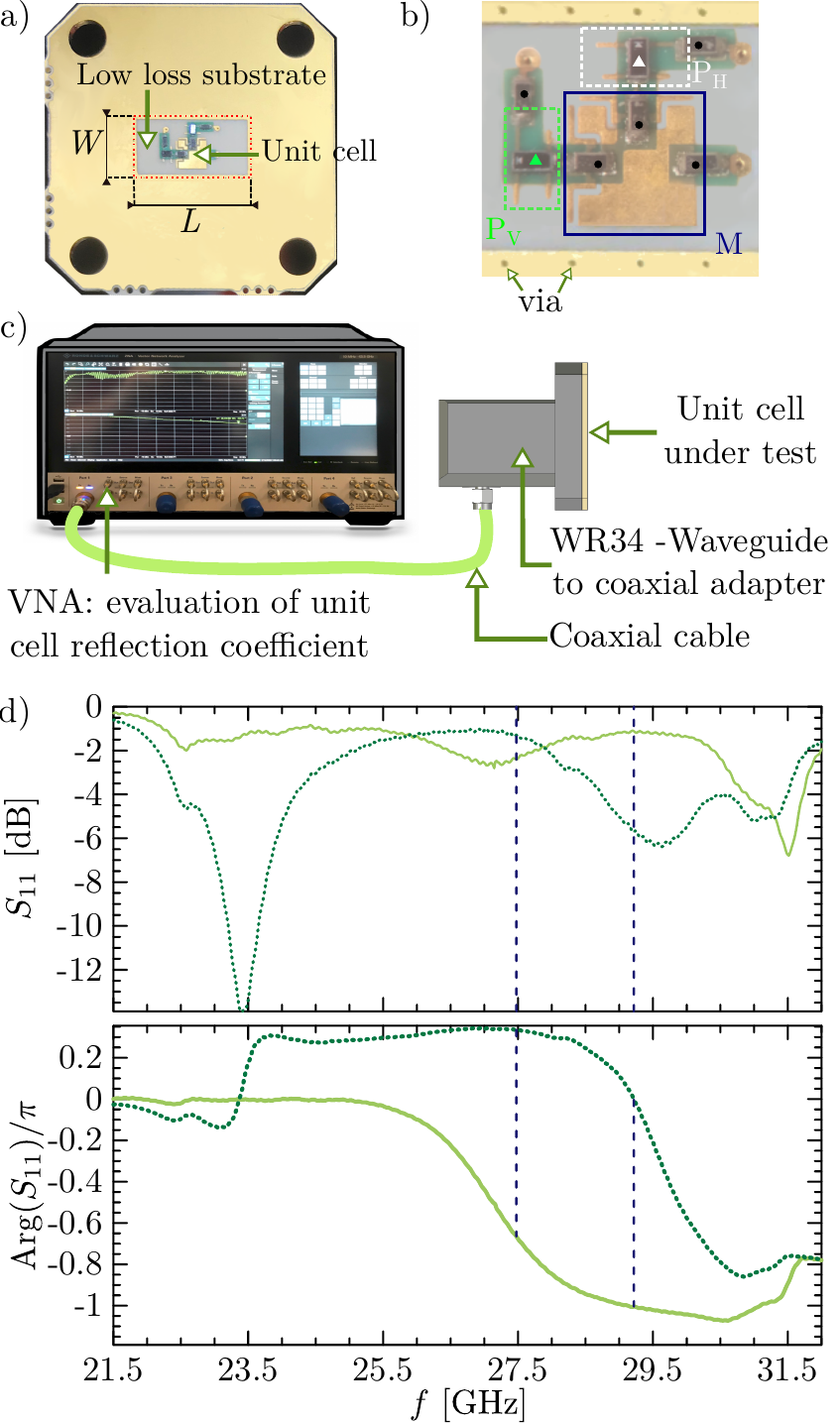}}
	\caption{\label{sec1:fig3}a) Pixel fabricated for the waveguide characterization. The big copper area around is intended to form electrical contact with the waveguide flange. b) Magnified photo of the fabricated pixel. The vias around provide isolation of the field inside waveguide.  $M$, $P_H$ and $P_V$ are main patch, horizontal and vertical parasitic resonator respectively. Dots and triangles tag respectively chokes and pin diodes. c) The setup of the experiment. d) Measured reflection from the single pixel inside the rectangular WR-34 waveguide. Solid line represents OFF-state and doted line represents ON-state of the pixel. Vertical dashed lines correspond to frequency where actual $\pi$-phase shift occurred showing the operating frequency range of the pixel.}
\end{figure}

To experimentally verify the properties of the designed pixel a set of pixels with slightly variable dimensions of resonators was fabricated. The WR-34 rectangular waveguide was used to characterize reflection properties of the single pixel. The pixel size was adjusted to fit the waveguide aperture (W$\times$L = 0.34$\times$0.17 inches). In order to avoid leakage of the field through the pixel substrate and provide isolation of the field inside the waveguide, a periodic via structure was formed at the edge of the pixel acting as an artificial electric wall. The bias network vias are connected with the pins on the back side of the pixel for the power supply connection. In order to place electronics on the back side of pixel there is a FR-4 substrate placed under the radiofrequency grounding layer (RF-ground). That is, single pixels consists of 2 layers of substrate separated by the ground layer.     
The pixel is tightly attached to the flange of the WR-34 waveguide in order to provide proper electrical connection with the waveguide. The waveguide (which is also the waveguide to coaxial adapter) is connected to a Rohde~\&~Schwarz  ZVA-40 vector network analyzer . 
In order to provide reflection state control, the electrical pins of the diodes, placed on the back side of the pixel, were connected to the constant voltage power source. The OFF-state of the diode was provided by the 0-voltage applied to the anode of the diode. The ON-state was provided by the -5V voltage applied to the cathode of the diode.  
The measured reflection properties of the single pixel inside a rectangular waveguide is shown on Fig.~\ref{sec1:fig3}. We see that we can indeed control the phase difference of the reflection coefficient by changing the position of the main patch resonance coupled with the resonance of parasitic resonator. In the 2~GHz frequency range between 27.5~GHz and 29.5~GHz the phase difference between ON and OFF sates is between $144^\circ$ and  $180^\circ$. The reflection amplitude in this range doesn't drop below -6~dB and average value in the frequency range bet wen 2 states is equal to -3~dB reaching -1~dB for some of the frequencies. It is worth noting that during the operation of the RIS, the average number of pixels in ON and OFF state are approximately equal. 
The difference between the simulated and the experimental results can be explained by the following reasons.  First there is the way we simulate the properties of the lumped components (PIN diodes, chokes). The equivalent RLC-model is used to describe the diodes and chokes properties and the parameters of the model may differ from the properties of real components. Secondly, in the simulation model,  we don't consider the effect of the FR4 layer placed behind the RF-ground. 

\subsection{Reconfigurable intelligent  surface fabrication}
 \begin{figure}
	\centerline{\includegraphics[width=\columnwidth]{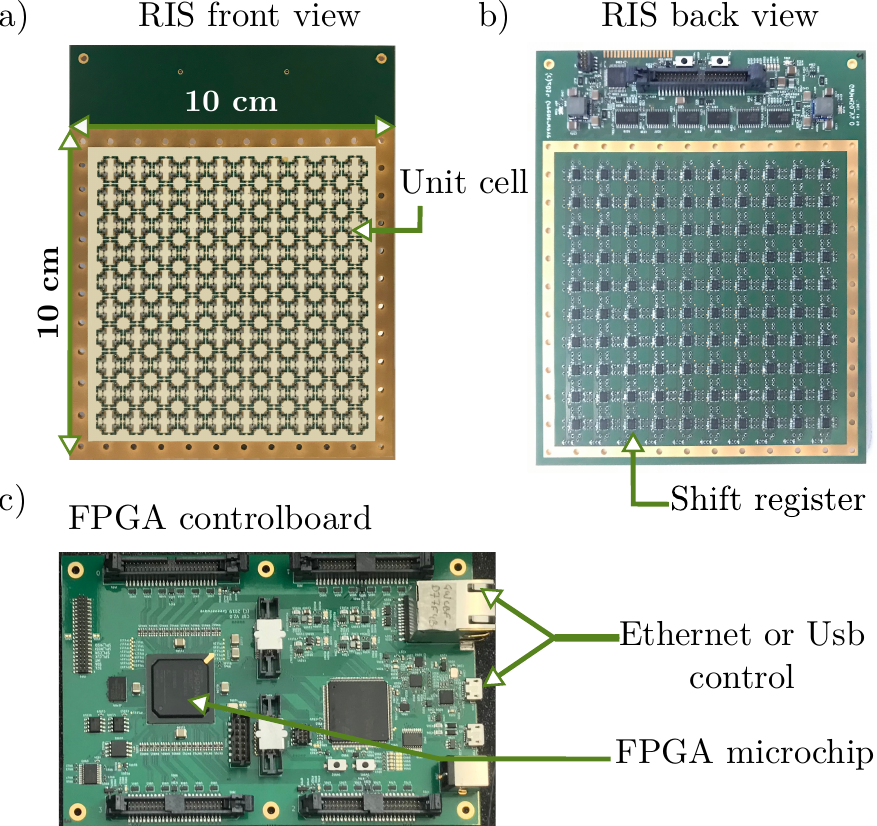}}
	\caption{\label{sec1:fig:metasurface} a) Front and b) back side of the fabricated $10~\times~10~\textrm{cm}^2$ RIS. c) RIS control board with different interfaces. }
\end{figure}
After the validation of the properties of 30 different pixels inside the waveguide the geometry which provides lower dissipation and better phase shift is selected. The selected pixel is used for fabricating the metasurface acting as a a reconfigurable intelligent surface (RIS).
 We fabricate a  $10~\times~10~\textrm{cm}^2$ RIS -- approximately $10\times10 \lambda$ ($\lambda$ is a wavelengths at 30~GHz)  -- made up of  400 unit cells periodically placed every $\lambda/2=0.5$~cm on a rectangular lattice of $20\times20$ pixel as shown by Fig.~\ref{sec1:fig:metasurface}.a). Each unit cells  having to control independently both components of the reflected EM field, an overall of 400 diodes  to control vertical polarisation and 400 diodes to control horizontal polarization are used. More technically speaking, our  RIS consists in a 6 layer PCB. The first layer is made from a low loss substrate, namely the METEORWAVE 8000 from AGC, and supports the 400 unit cells (see front view in Fig~\ref{sec1:fig:metasurface}.a)).  The 5 remaining layers are made from FR4 substrate. The last layer shown in  Fig~\ref{sec1:fig:metasurface}.b), notably supports the electronic components that allow us to control the states of the diodes of the pixels on the first layer. This control is operated by a $10 \times 10$  matrix of shift registers, each of them managing the states of 8 PIN diodes associated to the horizontal and vertical polarization states of 4 pixels.
In order to provide control of each pixel of the RIS, a FPGA control board has been designed and programmed (see Fig~\ref{sec1:fig:metasurface}.c)). This control board has  4 ports to connect 4 independent RIS. The control board is connected to the desktop through LAN or USB interface and is controlled with a Python or MATLAB code. The developed software allows to independently change the state of each diode on the RIS providing the required reflection properties across the whole RIS. PIN diodes are thus connected to the control board through the network of shift registers.
 We should point out that  the assembly composed by the metasurface  and the FPGA control board shown in Fig~\ref{sec1:fig:metasurface}   is a very low power consuming RIS. Indeed, it consumes in average no more than 8 Watts, with a potential  peak power of 16 Watts if all diodes are in ON-state. This power consumption can even be more reduced by tuning of the bias voltage applied to diode. In practice, in more recent version of our RIS, we have been able to reduce this power consumption by a factor of two.  A last interesting propriety of our RIS is its switching state rate.  While remotely controlling from a desktop, the latter  lies in a kHz range. If the pattern of the PIN diode states is stored locally on the board, update rate reaches nearly 100 kHz. Current maximum speed is function of the shift register chain organization. In the following, we will know see how to use this RIS for beamforming applications.

\section{Analytical model of binary phase RIS}
\begin{figure}[b]
	\centerline{\includegraphics[width=2.5in]{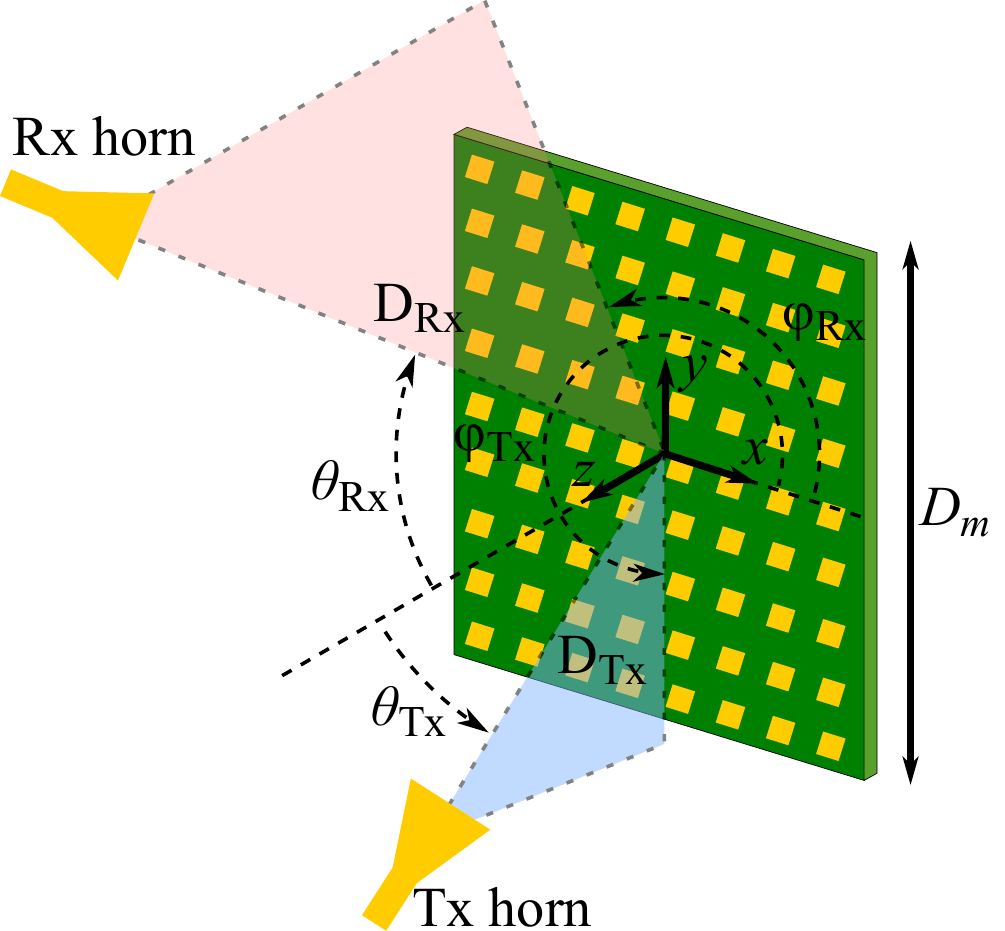}}
	\caption{Schematics of the mutual positions of the RIS, receiving (Rx) and transmitting (Tx) horn antennas in the spherical coordinates system. 
		\label{sec3:fig1}}
\end{figure}
\begin{figure*}[h]
	\centerline{\includegraphics[width=7in]{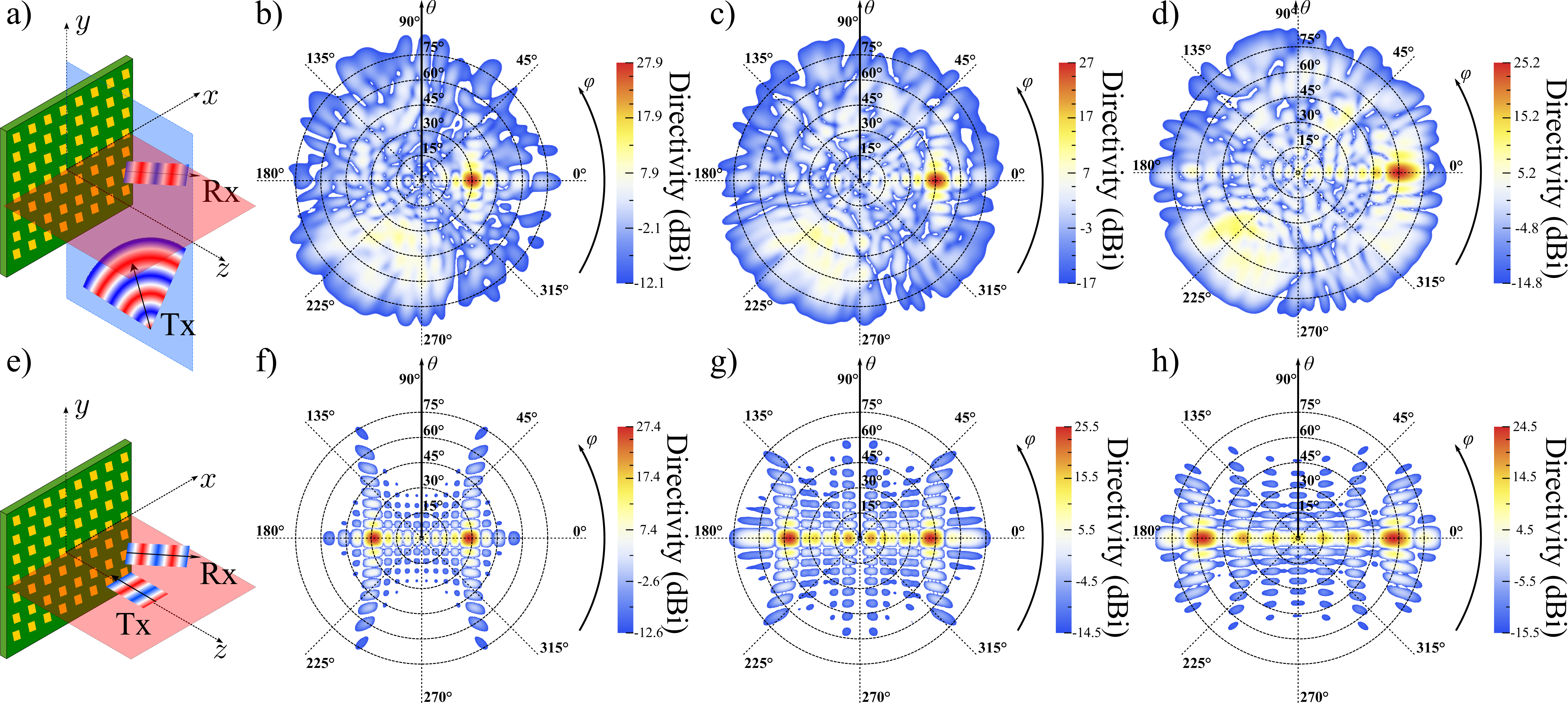}}
	\caption{(a) Illustration of a spherical wave impinging the RIS and a plane wave being reflected.
		(b)--(d) Results of the analytical model for the far-field radiation pattern created by the RIS for different configurations defined by steering directions: $\theta_{Rx}=30^\circ, 45^\circ$ and $60^\circ$ are shown ($\varphi_{Rx}$ is $0^\circ$).
		The Tx antenna is at $D_{Tx}=145$ mm, $\theta_{Tx}=45^\circ$ and $\varphi_{Tx}=270^\circ$.
		(e) Illustration of a plane wave impinging the RIS and a plane wave being reflected.
		(f)--(h) Results of the analytical model for far-field radiation patterns created by the RIS for different configurations defined by steering directions: $\theta_{Rx}= 30^\circ, 45^\circ$ and $60^\circ$ are shown ($\varphi_{Rx}$ is $0^\circ$).
		\label{sec3:fig2}}
\end{figure*}
In this section we derive an analytical model of binary phase RIS. This model is then used to investigate different  physical scenarios of a wave impinging the RIS, underlining  the peculiarities of   binary phase shifting metasurfaces and explaining their mechanism. The reflection of the electromagnetic wave impinging the RIS can be controlled at will by judiciously switching PIN diodes.
As it was shown in the previous section, states of PIN diodes determine the phase of the local reflection coefficient along the RIS.
The distribution of the phase $\phi_{nm}$, where $n$ and $m$ number pixels in the array, is established according to the following equation~\cite{Hum2014_review}:
\begin{equation}
    \label{eq:phase_grad}
    \phi_{nm} = \phi_r(x_n,y_m) - \phi_i(x_n,y_m),
\end{equation}
where $\phi_i(x_n,y_m)$ and $\phi_r(x_n,y_m)$ are the phase distributions created by the incident and (desired) reflected waves at the position $x_n$, $y_m$ of the $nm$th pixel.
The incident wave is created by the transmitting (Tx) antenna illuminating the RIS.
Since PIN diodes allow one to switch a pixel between two phase-states ($0$ and $\pi$) for each polarization, the phase $\phi_{nm}$ is set to $0$ when $-\pi/2\leq \phi_r(x_n,y_m) - \phi_i(x_n,y_m) \leq \pi/2$ and to $\pi$ otherwise.
The radiation pattern $E(\theta,\varphi)$ created by the RIS set to a given configuration can be approximated as the radiation pattern created by the array of patch-like antennas and reads ~\cite{yang2016programmable}
\begin{eqnarray}
    \label{eq:rad_pattern}
    &&E_r(\theta,\varphi) = \cos(\theta) \sum_{n,m = 1}^N \Gamma_{nm} E_i(x_n,y_m) \cos(\theta_{nm})\nonumber\\
    &&\times\exp(-jk\sin(\theta)[x_n\cos(\varphi)+y_m\sin(\varphi)]),
\end{eqnarray}
where $\cos(\theta)$ represents the radiation pattern of each individual antenna in the array,
$\Gamma_{nm}$ is the local reflection coefficient from the $nm$th pixel, 
$E_i(x_n,y_m)$ is the electric field of the incident wave at the $nm$th pixel,
$\theta_{nm}$ is the $\theta$-angle at which $nm$th pixel sees the Tx antenna, see Fig.~\ref{sec3:fig1}.

In what follows we focus on two practically important examples when: (i) Near-field configuration, \textit{i.e} the Tx antenna is placed at a distance to the RIS comparable to its size $D_m$  and (ii) Far-field configuration, \textit{i.e}   the Tx antenna is far away from the RIS such that the incident wave can be considered  as a plane wave. The receiving (Rx) horn antenna is always assumed to be far away from the RIS, i.e. $D_{Rx}\gg D_m$.
\subsection{Tx antenna is close to RIS: Near-field configuration}
The first scenario resembles a classical reflectarray antenna configuration.
The incident wave radiated by the Tx (horn) antenna is modeled as a spherical wave $\exp(-j k \sqrt{x^2+y^2+z^2})/\sqrt{x^2+y^2+z^2}$, 
although more elaborated models can be used to take into account such parameters of the antenna as its directivity. 
The reflected wave is set to be a plane wave $\exp(-jk\textbf{r}\textbf{n})$, which propagates towards the Rx antenna in the direction $\theta_{Rx}$, $\varphi_{Rx}$ defined by the vector $\textbf{n}$.
Under this configuration Eq.~\eqref{eq:phase_grad} takes the following form:
\begin{eqnarray}
    \label{eq:phase_NF_FF}
    &&\phi_{nm} = - k\sin(\theta_{Rx})[x_n\cos(\varphi_{Rx}) + y_m\sin(\varphi_{Rx})]
    \nonumber\\
    &&+ k \sqrt{(x_n-x_{Tx})^2+(y_n-y_{Tx})^2+z_{Tx}^2},
\end{eqnarray}
where $z_{Tx}=D_{Tx}\cos(\theta_{Tx})$, $x_{Tx}=D_{Tx}\sin(\theta_{Tx})\cos(\varphi_{Tx})$, $y_{Tx}=D_{Tx}\sin(\theta_{Tx})\sin(\varphi_{Tx})$.

Fig~\ref{sec3:fig2} (a)--(d) demonstrates different numerical examples of beam-forming with $10\lambda\times 10\lambda$ large RIS. 
The configuration of the RIS is set  according to Eq.~\eqref{eq:phase_NF_FF} and the resulting radiation pattern is calculated with Eq.~\eqref{eq:rad_pattern}.
The colorbars in panels of Fig~\ref{sec3:fig2}.b)--d) indicate the directivity in each particular example and help to understand how well the incident wave is steered by RIS in the direction of a receiving antenna.
The directivity is maximum in Fig.~\ref{sec3:fig2}.b) corresponding to $\theta_{Rx}=30^\circ$ steering angle and reaching $27.9$ dBi value. 
When increasing the steering angle to $\theta_{Rx}=60^\circ$, the directivity drops to $25.2$ dBi.
These values can be compared to the maximum possible directivity of approximately $30.5$ dBi, which is created by the uniform aperture of the same area as the RIS ($A=100\lambda^2$) at $\theta_{Rx}=0^\circ$ and calculated as $4\pi A/\lambda^2$ \cite{balanis2016antenna}.
The directivity of the RIS decreases from the maximum value due to the following factors: (i) the steering angle decreases the physical aperture of RIS, (ii) only two phase-states are available for a binary RIS, which increases the level of undesired secondary lobes,
(iii) the incident spherical wave does not illuminate the RIS uniformly.
Out of these three factors, only the last one is responsible for the reduced directivity in the near-field configuration.
Indeed,
the aperture size sets a fundamental limit on the directivity and only by increasing the RIS area this limit can be increased.
Even though the considered RIS is binary, the level of secondary lobes, in the near-field configuration, remains low even for such large steering angles as $60^\circ$.
However, the near field configuration also implies that a RIS cannot be effectively illuminated and this  decreases the aperture efficiency and thus the effective size of the physical aperture.

\subsection{Tx antenna is far from RIS:  Far-field configuration}

In the second case, the Tx antenna is assumed to be further away from the RIS such that $D_{Tx}\gg D_m$, when the impinging wave can be well approximated as a plane wave.
The phase profile along the RIS is thus calculated by means of the following formula
\begin{eqnarray}
    \label{eq:phase_FF_FF}
    &&\phi_{nm} = - k\sin(\theta_{Rx})[x_n\cos(\varphi_{Rx}) + y_m\sin(\varphi_{Rx})]
    \nonumber\\
    &&+ k\sin(\theta_{Tx})[x_n\cos(\varphi_{Tx}) + y_m\sin(\varphi_{Tx})].
\end{eqnarray}
It leads to a periodic phase profile which cannot be well-approximated by only two phase states.
The binary RIS put in a periodic configuration does not allow one to control the far-field radiation pattern completely and a very strong undesired lobe appears in the addition to the main lobe.
This effect is well demonstrated by Figs.~\ref{sec3:fig2}.e)--h) where a normally incident plane wave is split in two symmetrical beams according to periods of set configurations.
It is curious to note that although the strong secondary lobe reduces the directivity, the values calculated in the three examples are very close to those in the corresponding near-field configurations (see Figs.~\ref{sec3:fig2}.b)--d)).
It can be explained by the fact that the plane-wave illumination creates the maximum aperture efficiency of $1$, that cannot be achieved with a Tx antenna close to a RIS.
To conclude this section, the analytical model allows one to find out in advance a configuration of RIS for a given functionality and estimate its efficiency in terms of the directivity.
More elaborate algorithm can be implemented on the basis of Eq.~\eqref{eq:rad_pattern} to control with RIS multiple or shaped beams~\cite{yang2016programmable}.
An analytical configuration can also be a starting point for an experimental optimization procedure with an implemented feedback mechanism as the one proposed in the following experimental section.
\section{Experimental results}
\begin{figure}[t]
	\centerline{\includegraphics[width=\columnwidth]{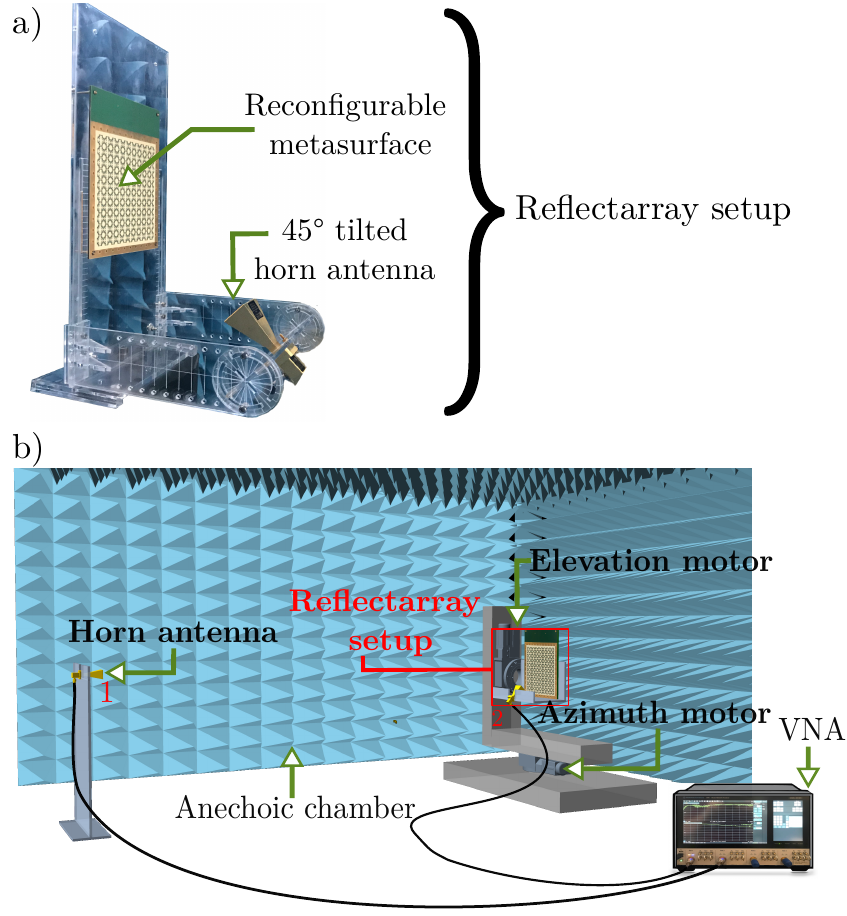}}
	\caption{\label{sec4:figreflect_setup} a) Photograph of the experimental reflectarray setup with our  $10$~cm by $10$~cm RIS.  b) Experimental setup for measuring the far field radiation pattern of a).}
\end{figure}
\begin{figure}[t]
	\centerline{\includegraphics[width=3.5in]{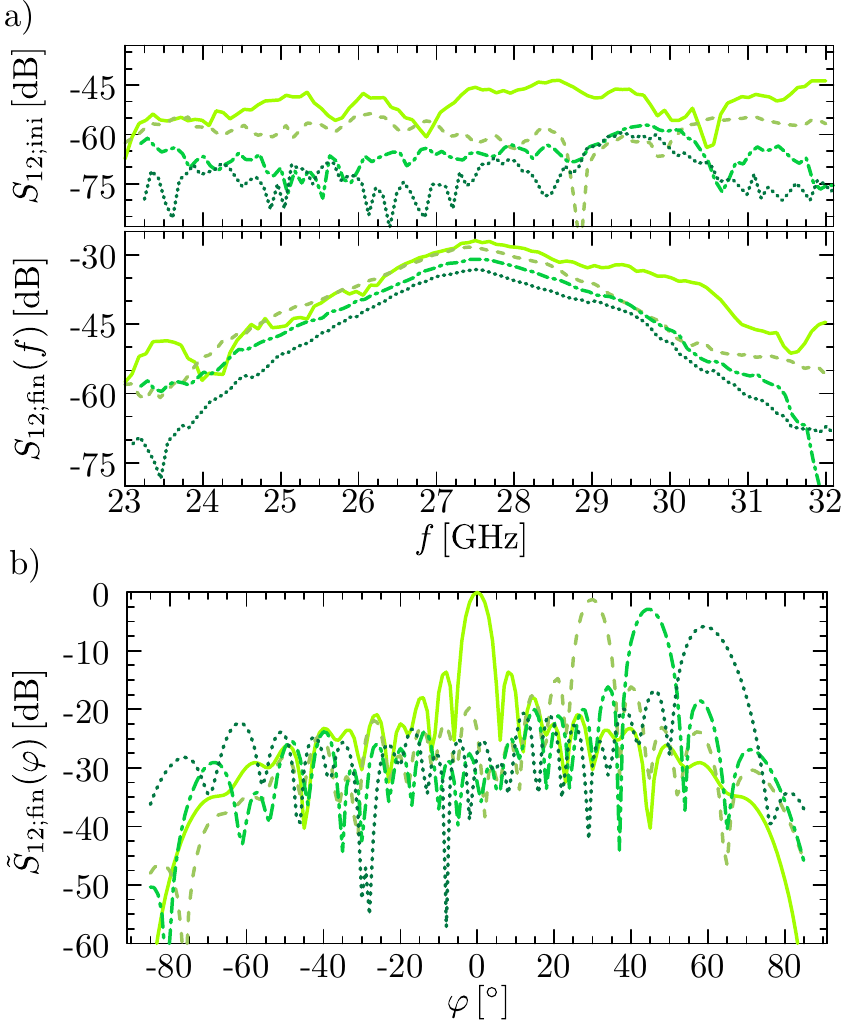}}
	\caption{
		a) Experimentally measured frequency response of the reflectarray setup of Fig~\ref{sec4:figreflect_setup} in the steering directions $\theta_{Rx}=0^\circ$ (continuous curves), $\theta_{Rx}=30^\circ$ (dashed  curves), $\theta_{Rx}=45^\circ$ (dashed-dotted curves) and $theta_{Rx}=60^\circ$ (dotted curve) ($\varphi_{Rx}$ is $0^\circ$) for the initial RIS configuration (top figure) and for the optimized configurations (bottom figure) . b) Experimentally measured normalized far-field radiation patterns for the same steering angles and for the optimized configurations .
		\label{sec4:fig2}}
\end{figure}
In this section we investigate experimentally the  two possible physical situations mentioned in the above section and corresponding to RIS placed respectively in the near field (Fig~\ref{sec3:fig2}.a)) or in the far field (Fig~\ref{sec3:fig2}.e)) of the Tx antenna.  
\subsection{Reflectarray setup: RIS in Near-field configuration}
To study the first situation, a horn antenna is placed $145$ mm away from our mmWave RIS  and  tilted $45^\textrm{\degree}$ in odrer to reduce shading by the feed horn. The overall set-up is mounted on a mmWave transparent plastic holder as shown in Fig~\ref{sec4:figreflect_setup}.a). In order to measure the radiation pattern of the thus-obtained reflectarray set-up, the latter is fixed on an Elevation-Azimuth rotation stage and installed in an anechoic chamber  in front of a fixed horn Tx antenna 1 as depicted in Fig~\ref{sec4:figreflect_setup}.b).  Then, for different relative angular position ($\varphi_0,\theta_0$) between the reflectarray setup and the fixed RX horn antenna 1, starting from an initial configuration of the RIS  corresponding with all pixel in on state, we optimize the configuration of the latter  in order to increase the amplitude of the scattering parameter measured by the VNA at $27.5 $~GHz between the Rx horn antenna 2 on the reflectarray setup and the Tx antenna. The optimization scheme we use is a straightforward iterative one comprising measuring the transmission amplitude $\abs{S_{12}}$  for each pixel for  its four possible combinations of  PIN diodes states and and selecting the state of the pixel where $\abs{S_{12}}$ is maximized. This operation is repeated for all pixels. This optimization scheme is inspired from one commonly used in adaptive optics \cite{Bridges74} and has been successfully implemented  during the last decade in numerous wavefront shaping applications in complex media dealing equally with optical  \cite{Vellekoop2007,Mosk2012} , electromagnetic waves \cite{Kaina2014,Hougne} and even acoustic waves \cite{Ma2018}. Then, after optimization, we measure the far field radiation pattern of the optimized configuration.
In Fig~\ref{sec4:fig2}.a), the top and bottom figure respectively correspond to 
the frequency responses for the initial configuration and the optimized one 
for four  different azimuth angles $\varphi_0=0^\textrm{\degree},30^\textrm{\degree},45^\textrm{\degree}$  and $60^\textrm{\degree}$ and a constant elevation angle $\theta_0=0^\textrm{\degree}$. For each angles, the optimized transmissions  are at the optimized frequency about $30$~dB above the initial ones. Fig~\ref{sec4:fig2}.~b) shows four radiation patterns associated with optimized configuration for elevation angle $\theta_0=0^\textrm{\degree}$ and azimuth angles $\varphi_0=0^\textrm{\degree}$ (continuous line) , $\varphi_0=30^\textrm{\degree}$  (dashed line), $\varphi_0=45^\textrm{\degree}$  (dashed-dotted line)  and $\varphi_0=60^\textrm{\degree}$ (dotted line). This figure is in good agreement with the numerical model shown above and demonstrates experimentally the ability of our millimeter-wave RIS to steer a beam even at $60^\textrm{\degree}$. At that point, we would like to draw the reader's attention to the fact that, regardless of the steering angle, the RIS configuration obtained through the optimization process is very similar to the one we would predict with the numerical model describes in the previous section. However the optimized configuration generally leads to a field which is about 2.5dB higher in absolute value in the direction of the main beam. This is  due to the fact that optimization process is able to deal with imperfections in the manufacturing of the RIS (not all pixels are perfectly equivalent), and probably also with a  slight misalignment in the experimental setup. We can also note that the first sidelobe levels are about $-13$~dB as expected for rectangular  aperture. And finally, due to the decrease in the effective size of the geometric aperture as the optimization angles increase, the beamwidth and gain naturally increases and decreases respectively for higher optimization angles. Furthermore,  with the setup of  Fig~\ref{sec4:figreflect_setup}, we have checked that the beam  can been steered in a range of $120^\textrm{\degree}$ with a precision of  $0.2^\textrm{\degree}$.

\subsection{
RIS in access-point extender configuration
}
\begin{figure}[h]
	\centerline{\includegraphics[width=3.5in]{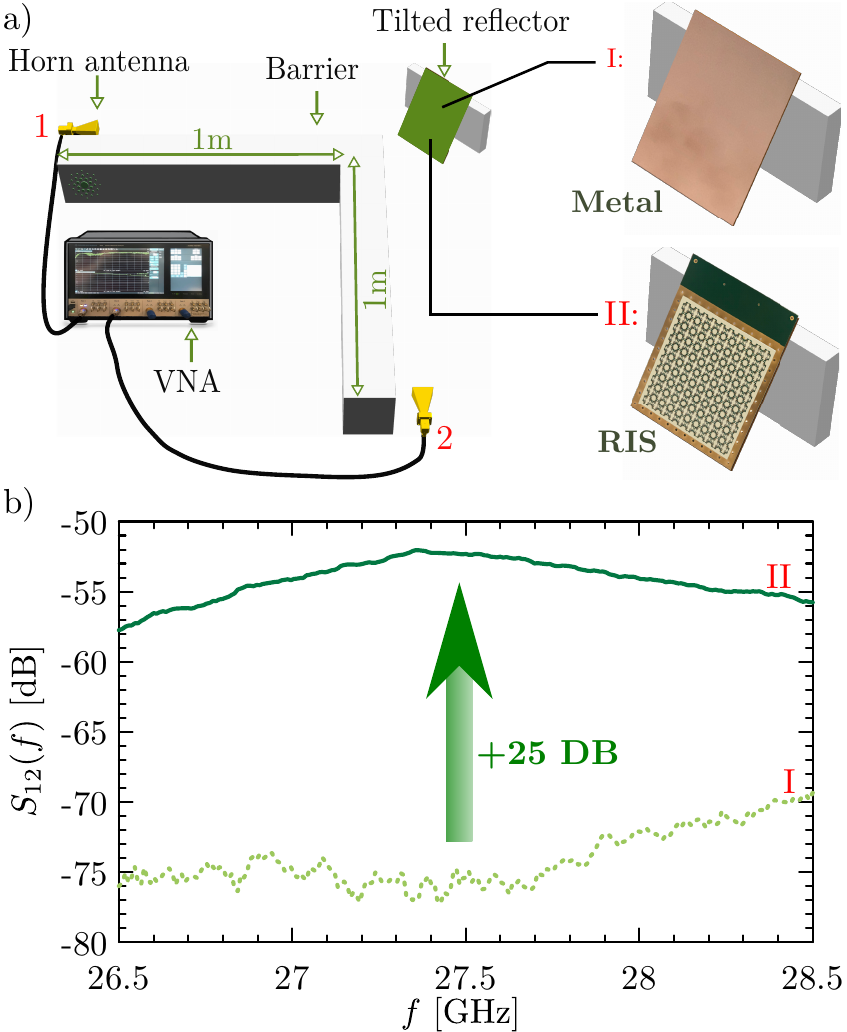}}
	\caption{
		(a) Experimental setup emulating a wireless transmission of mmWaves  between two horn antennas connected to a VNA. Line of sight between the receiver and the transmitter is suppress by a right angle barrier.  A tilted reflector  face the corner of the barrier.	In scenario I, the reflector is a metallic plate which is replace by a RIS in scenario II. b) Comparison of transmission parameter measured in scenario I (light green dotted curve) and after optimization of  the RIS configuration in scenario II (dark green continuous curve ) \label{sec4:fig3}}
\end{figure}
The second physical situation we are experimentally addressing is the RIS placed in the far-field of both Rx and Tx antenna as illustrated in  Fig~\ref{sec3:fig2}.e). 
In order to play with the RIS in a far field configuration, we propose the  wireless communication scenario at mmWave depicted in Fig;~\ref{sec4:fig3}.a). The latter consists in a wireless transmission of mmWaves  between two horn antennas connected to a VNA while we avoid any line of sight between the receiver and the transmitter by placing each antenna at both extremity of a right angle barrier (emulating for example the intersection between two corridors). Each arm of the barrier is $1$~m long.  The overall setup is installed  outside the anechoic chamber to be closer to an actual wireless communication scenario. 
Strictly speaking, we are not in the far-field of the RIS in antenna terms (by the factor of 2), but the Tx and Rx antenna do not have a line of sight and are one order of magnitude further away from the RIS in comparison to the previous experiment. However, we can still  consider the RIS to be in the far-field zone of a Rx or a Tx antenna if  the field radiated by either of these antennas can be well-approximated by a plane wave at the position of the RIS. As we use in our experiments simple horn antennas (having around 17 dBi gain), the far-field zone for these antennas is around 15 cm.

To demonstrate the advantage of  using a RIS in this kind of problem,  we measure the transmission between the antennas in two different situations. In the situation I, a tilted metallic plate is facing the corner of the barrier. The light green dotted curve in Fig.~\ref{sec4:fig3}.b)  corresponds to the obtained transmission coefficient  in a frequency range  of $2$~GHz around $27.5$~GHz. In this situation the transmission between the antennas is very low. In the situation II, the metallic plate is replaced by a RIS which is also tilted. The relative position between the horn antennas 1 and 2 and the RIS not being evaluated, we use the same iterative optimization scheme describe above to find an  optimized configuration that  increase the mean value of the transmission coefficient $\langle \abs{S_{12}(f) }\rangle_f`$ in a frequency window of $250$~MHz around  $27.5$~GHz. The frequency response associated to the optimized configuration  is plotted in  Fig.~\ref{sec4:fig3}.b (dark green continuous curve). Around the optimized frequency windows, we measure  a $25$~dB gain over the received energy. We show also that the  energy gain is valid well beyond the $250$~MHz frequency window where we did the optimization. This is consistent with the bandwidth of the metasurface. Thus, we have proposed an experimental proof of wireless transmission of mmWaves where a normally absent line of sight is restored thanks to our RIS acting as a highly efficient and  and low-power consumption access point extender. It is worth noting that here  a $25$~dB gain is  obtained merely on  cost of the power dissipated by the PIN diodes, namely 8 Watts in average for the whole metasurface (reduced to 4 Watts in our last version of metasurface).

As a remark, in this experiment we used a metal plate as a reference, but RIS in a random configuration can also be used as a benchmark. However, the result will depend on the particular random configuration and an averaging over a large ensemble of configurations has to be performed.

\begin{table}[tb]
    \centering
        \caption{RIS specifications table.}
    \label{tab1}
    \begin{tabular}{|c|c|}
    \hline
    Aperture size & $10$ cm $\times$ $10$ cm\\ \hline
    Number of elements  & 400 \\ \hline
    Polarization & dual linear \\ \hline    
    Element's spacing & $\lambda/2$ \\  \hline
    Operating frequency range  & $[27.5$ GHz, $29.5$GHz$]$\\  \hline
    Instantaneous bandwidth & 500 MHz\\ \hline
    Directivity  & $22.5$ dBi\\  \hline
    Scan range (El., Az.) & $\pm 60^\circ$\\ \hline
    Switching rate & 100 kHz\\
    \hline
    \end{tabular}

\end{table}

\section{Conclusion}
In this paper, we introduced a design of the binary RIS operating in a transmitting spectrum of the Ka-band, key performance parameters can be found in Tab.~\ref{tab1}. The design of the single pixel of the RIS provides low level of reflection wave dissipation, near-$\pi$ value of the phase difference and wide instantaneous bandwidth.
We showed that the RIS can be used in two different configurations: as a general reflection array illuminated by a closely positioned horn antenna and as highly efficient and low power consumption access point extender. In the last case, the RIS provides an amplification of the signal between transmitter and receiver in a scenario where originally there is no line of sight between them.   
In the future work we plan to demonstrate the functionality of the developed RIS as a part of communication system using Software-Defined-Radio, especially regarding robustness of the link with respect to the power noise potentially introduced by the RIS.

\bibliographystyle{ieeetr}

\begin{thebibliography}{10}
	
	\bibitem{Foschini1998}
	G.~J. Foschini and M.~J. Gans, ``{On Limits of Wireless Communications in a
		Fading Environment when Using Multiple Antennas},'' {\em Wirel. Pers.
		Commun.}, vol.~6, no.~3, pp.~311--335, 1998.
	
	\bibitem{paulraj2003introduction}
	A.~Paulraj, A.~Rohit, A.~Eringen, R.~Nabar, D.~Gore, and C.~U. Press, {\em
		Introduction to Space-Time Wireless Communications}.
	\newblock Cambridge University Press, 2003.
	
	\bibitem{Tarokh1998}
	V.~Tarokh, N.~Seshadri, and A.~Calderbank, ``{Space-time codes for high data
		rate wireless communication: performance criterion and code construction},''
	{\em IEEE Trans. Inf. Theory}, vol.~44, pp.~744--765, mar 1998.
	
	\bibitem{Alamouti1998}
	S.~Alamouti, ``{A simple transmit diversity technique for wireless
		communications},'' {\em IEEE J. Sel. Areas Commun.}, vol.~16, no.~8,
	pp.~1451--1458, 1998.
	
	\bibitem{Marzetta2013}
	T.~L. {Marzetta}, G.~{Caire}, M.~{Debbah}, I.~{Chih-Lin}, and S.~K. {Mohammed},
	``Special issue on massive mimo,'' {\em Journal of Communications and
		Networks}, vol.~15, no.~4, pp.~333--337, 2013.
	
	\bibitem{Larsson2014}
	E.~G. {Larsson}, O.~{Edfors}, F.~{Tufvesson}, and T.~L. {Marzetta}, ``Massive
	mimo for next generation wireless systems,'' {\em IEEE Communications
		Magazine}, vol.~52, no.~2, pp.~186--195, 2014.
	
	\bibitem{Lu2014}
	L.~{Lu}, G.~Y. {Li}, A.~L. {Swindlehurst}, A.~{Ashikhmin}, and R.~{Zhang}, ``An
	overview of massive mimo: Benefits and challenges,'' {\em IEEE Journal of
		Selected Topics in Signal Processing}, vol.~8, no.~5, pp.~742--758, 2014.
	
	\bibitem{Bjornson2015}
	E.~{Björnson}, L.~{Sanguinetti}, J.~{Hoydis}, and M.~{Debbah}, ``Optimal
	design of energy-efficient multi-user mimo systems: Is massive mimo the
	answer?,'' {\em IEEE Transactions on Wireless Communications}, vol.~14,
	no.~6, pp.~3059--3075, 2015.
	
	\bibitem{Bjornson2016}
	E.~{Björnson}, E.~G. {Larsson}, and T.~L. {Marzetta}, ``Massive mimo: ten
	myths and one critical question,'' {\em IEEE Communications Magazine},
	vol.~54, no.~2, pp.~114--123, 2016.
	
	\bibitem{Subrt1}
	L.~Subrt, D.~Grace, and P.~Pechac, ``Controlling the short-range propagation
	environment using active frequency selective surfaces,'' {\em
		Radioengineering}, vol.~19, 12 2010.
	
	\bibitem{Subrt2}
	L.~{Subrt} and P.~{Pechac}, ``Intelligent walls as autonomous parts of smart
	indoor environments,'' {\em IET Communications}, vol.~6, no.~8,
	pp.~1004--1010, 2012.
	
	\bibitem{Kaina2014}
	N.~Kaina, M.~Dupr{\'e}, G.~Lerosey, and M.~Fink, ``Shaping complex microwave
	fields in reverberating media with binary tunable metasurfaces,'' {\em
		Scientific Reports}, vol.~4, p.~6693, Oct 2014.
	
	\bibitem{Dupre2015}
	M.~Dupr{\'{e}}, P.~del Hougne, M.~Fink, F.~Lemoult, and G.~Lerosey,
	``{Wave-Field Shaping in Cavities: Waves Trapped in a Box with Controllable
		Boundaries},'' {\em Phys. Rev. Lett.}, vol.~115, p.~017701, jul 2015.
	
	\bibitem{X.Tan}
	X.~{Tan}, Z.~{Sun}, J.~M. {Jornet}, and D.~{Pados}, ``Increasing indoor
	spectrum sharing capacity using smart reflect-array,'' in {\em 2016 IEEE
		International Conference on Communications (ICC)}, pp.~1--6, 2016.
	
	\bibitem{Welkie}
	A.~Welkie, L.~Shangguan, J.~Gummeson, W.~Hu, and K.~Jamieson, ``Programmable
	radio environments for smart spaces,'' in {\em Proceedings of the 16th ACM
		Workshop on Hot Topics in Networks}, HotNets-XVI, (New York, NY, USA),
	p.~36–42, Association for Computing Machinery, 2017.
	
	\bibitem{Hougne}
	P.~del Hougne, M.~Fink, and G.~Lerosey, ``Optimally diverse communication
	channels in disordered environments with tuned randomness,'' {\em Nature
		Electronics}, vol.~2, pp.~36--41, Jan 2019.
	
	\bibitem{Liaskos_7}
	C.~Liaskos, S.~Nie, A.~Tsioliaridou, A.~Pitsillides, S.~Ioannidis, and
	I.~Akyildiz, ``Realizing wireless communication through software-defined
	hypersurface environments,'' in {\em 2018 IEEE 19th International Symposium
		on "A World of Wireless, Mobile and Multimedia Networks" (WoWMoM)}, (Los
	Alamitos, CA, USA), pp.~14--15, IEEE Computer Society, jun 2018.
	
	\bibitem{Liaskos_8}
	C.~{Liaskos}, S.~{Nie}, A.~{Tsioliaridou}, A.~{Pitsillides}, S.~{Ioannidis},
	and I.~{Akyildiz}, ``A new wireless communication paradigm through
	software-controlled metasurfaces,'' {\em IEEE Communications Magazine},
	vol.~56, no.~9, pp.~162--169, 2018.
	
	\bibitem{Renzo2019}
	M.~D. Renzo, M.~Debbah, D.-T. Phan-Huy, A.~Zappone, M.-S. Alouini, C.~Yuen,
	V.~Sciancalepore, G.~C. Alexandropoulos, J.~Hoydis, H.~Gacanin, J.~d. Rosny,
	A.~Bounceur, G.~Lerosey, and M.~Fink, ``Smart radio environments empowered by
	reconfigurable ai meta-surfaces: an idea whose time has come,'' {\em EURASIP
		Journal on Wireless Communications and Networking}, vol.~2019, p.~129, May
	2019.
	
	\bibitem{Basar2019}
	E.~{Basar}, M.~{Di Renzo}, J.~{De Rosny}, M.~{Debbah}, M.~{Alouini}, and
	R.~{Zhang}, ``Wireless communications through reconfigurable intelligent
	surfaces,'' {\em IEEE Access}, vol.~7, pp.~116753--116773, 2019.
	
	\bibitem{8741198}
	C.~{Huang}, A.~{Zappone}, G.~C. {Alexandropoulos}, M.~{Debbah}, and C.~{Yuen},
	``Reconfigurable intelligent surfaces for energy efficiency in wireless
	communication,'' {\em IEEE Transactions on Wireless Communications}, vol.~18,
	no.~8, pp.~4157--4170, 2019.
	
	\bibitem{DiRenzo2020}
	M.~{Di Renzo}, F.~{Habibi Danufane}, X.~{Xi}, J.~{de Rosny}, and
	S.~{Tretyakov}, ``Analytical modeling of the path-loss for reconfigurable
	intelligent surfaces – anomalous mirror or scatterer ?,'' in {\em 2020 IEEE
		21st International Workshop on Signal Processing Advances in Wireless
		Communications (SPAWC)}, pp.~1--5, 2020.
	
	\bibitem{Garcia2020}
	J.~C.~B. {Garcia}, A.~{Sibille}, and M.~{Kamoun}, ``Reconfigurable intelligent
	surfaces: Bridging the gap between scattering and reflection,'' {\em IEEE
		Journal on Selected Areas in Communications}, vol.~38, no.~11,
	pp.~2538--2547, 2020.
	
	\bibitem{Basar2020}
	E.~{Basar}, ``Reconfigurable intelligent surface-based index modulation: A new
	beyond mimo paradigm for 6g,'' {\em IEEE Transactions on Communications},
	vol.~68, no.~5, pp.~3187--3196, 2020.
	
	\bibitem{Nadeem2020}
	Q.~{Nadeem}, H.~{Alwazani}, A.~{Kammoun}, A.~{Chaaban}, M.~{Debbah}, and
	M.~{Alouini}, ``Intelligent reflecting surface-assisted multi-user miso
	communication: Channel estimation and beamforming design,'' {\em IEEE Open
		Journal of the Communications Society}, vol.~1, pp.~661--680, 2020.
	
	\bibitem{He2020}
	Z.~{He} and X.~{Yuan}, ``Cascaded channel estimation for large intelligent
	metasurface assisted massive mimo,'' {\em IEEE Wireless Communications
		Letters}, vol.~9, no.~2, pp.~210--214, 2020.
	
	\bibitem{Park2020}
	S.~Y. {Park} and D.~{In Kim}, ``Intelligent reflecting surface-aided
	phase-shift backscatter communication,'' in {\em 2020 14th International
		Conference on Ubiquitous Information Management and Communication (IMCOM)},
	pp.~1--5, 2020.
	
	\bibitem{Badiu2020}
	M.~{Badiu} and J.~P. {Coon}, ``Communication through a large reflecting surface
	with phase errors,'' {\em IEEE Wireless Communications Letters}, vol.~9,
	no.~2, pp.~184--188, 2020.
	
	\bibitem{Wu2020}
	Q.~{Wu} and R.~{Zhang}, ``Beamforming optimization for wireless network aided
	by intelligent reflecting surface with discrete phase shifts,'' {\em IEEE
		Transactions on Communications}, vol.~68, no.~3, pp.~1838--1851, 2020.
	
	\bibitem{Zappone2020}
	A.~{Zappone}, M.~{Di Renzo}, and M.~{Debbah}, ``Wireless networks design in the
	era of deep learning: Model-based, ai-based, or both?,'' {\em IEEE
		Transactions on Communications}, vol.~67, no.~10, pp.~7331--7376, 2019.
	
	\bibitem{Huang2019}
	C.~{Huang}, G.~C. {Alexandropoulos}, C.~{Yuen}, and M.~{Debbah}, ``Indoor
	signal focusing with deep learning designed reconfigurable intelligent
	surfaces,'' in {\em 2019 IEEE 20th International Workshop on Signal
		Processing Advances in Wireless Communications (SPAWC)}, pp.~1--5, 2019.
	
	\bibitem{DiRenzo2019}
	M.~{Di Renzo} and J.~Song, ``{Reflection probability in wireless networks with
		metasurface-coated environmental objects: an approach based on random spatial
		processes},'' {\em EURASIP J. Wirel. Commun. Netw.}, vol.~2019, p.~99, dec
	2019.
	
	\bibitem{Kaina:14}
	N.~Kaina, M.~Dupr\'{e}, M.~Fink, and G.~Lerosey, ``Hybridized resonances to
	design tunable binary phase metasurface unit cells,'' {\em Opt. Express},
	vol.~22, pp.~18881--18888, Aug 2014.
	
	\bibitem{5765450}
	H.~{Kamoda}, T.~{Iwasaki}, J.~{Tsumochi}, T.~{Kuki}, and O.~{Hashimoto},
	``60-ghz electronically reconfigurable large reflectarray using single-bit
	phase shifters,'' {\em IEEE Transactions on Antennas and Propagation},
	vol.~59, no.~7, pp.~2524--2531, 2011.
	
	\bibitem{7448838}
	H.~{Yang}, F.~{Yang}, S.~{Xu}, Y.~{Mao}, M.~{Li}, X.~{Cao}, and J.~{Gao}, ``A
	1-bit $10 \times 10$ reconfigurable reflectarray antenna: Design,
	optimization, and experiment,'' {\em IEEE Transactions on Antennas and
		Propagation}, vol.~64, no.~6, pp.~2246--2254, 2016.
	
	\bibitem{7902179}
	H.~{Yang}, F.~{Yang}, X.~{Cao}, S.~{Xu}, J.~{Gao}, X.~{Chen}, M.~{Li}, and
	T.~{Li}, ``A 1600-element dual-frequency electronically reconfigurable
	reflectarray at x/ku-band,'' {\em IEEE Transactions on Antennas and
		Propagation}, vol.~65, no.~6, pp.~3024--3032, 2017.
	
	\bibitem{meteorwave}
	V.~L. L. P.~M. AGC-Nelco High~Speed, ``{METEORWAVE 8000}.''
	\url{https://agc-nelco.com/products/meteorwave-8000//}.
	
	\bibitem{Hum2014_review}
	S.~V. Hum and J.~Perruisseau-Carrier, ``Reconfigurable reflectarrays and array
	lenses for dynamic antenna beam control: A review,'' {\em IEEE Transactions
		on Antennas and Propagation}, vol.~62, pp.~183--198, Jan 2014.
	
	\bibitem{yang2016programmable}
	H.~Yang, X.~Cao, F.~Yang, J.~Gao, S.~Xu, M.~Li, X.~Chen, Y.~Zhao, Y.~Zheng, and
	S.~Li, ``A programmable metasurface with dynamic polarization, scattering and
	focusing control,'' {\em Scientific reports}, vol.~6, p.~35692, 2016.
	
	\bibitem{balanis2016antenna}
	C.~A. Balanis, {\em Antenna theory: analysis and design}.
	\newblock John Wiley \& Sons, 2016.
	
	\bibitem{Bridges74}
	W.~B. Bridges, P.~T. Brunner, S.~P. Lazzara, T.~A. Nussmeier, T.~R. O'Meara,
	J.~A. Sanguinet, and W.~P. Brown, ``Coherent optical adaptive techniques,''
	{\em Appl. Opt.}, vol.~13, pp.~291--300, Feb 1974.
	
	\bibitem{Vellekoop2007}
	I.~M. Vellekoop and A.~P. Mosk, ``Focusing coherent light through opaque
	strongly scattering media,'' {\em Opt. Lett.}, vol.~32, pp.~2309--2311, Aug
	2007.
	
	\bibitem{Mosk2012}
	A.~P. Mosk, A.~Lagendijk, G.~Lerosey, and M.~Fink, ``{Controlling waves in
		space and time for imaging and focusing in complex media},'' {\em Nat.
		Photonics}, vol.~6, no.~5, pp.~283--292, 2012.
	
	\bibitem{Ma2018}
	G.~Ma, X.~Fan, P.~Sheng, and M.~Fink, ``{Shaping reverberating sound fields
		with an actively tunable metasurface},'' {\em Proc. Natl. Acad. Sci.},
	vol.~115, pp.~6638--6643, jun 2018.
	
\end{thebibliography}

\begin{IEEEbiographynophoto}
{Jean-Baptiste Gros} earned a PhD in Physics from Université de Nice in 2014. He is specialized in electromagnetic compatibility, wave chaos and random matrix theory approaches for modeling wave systems and wave propagation in complex media. After a postdoc in acoustics at LAUM (CNRS, Le Mans Université) and in electromagnetic at ISAE-SUPAERO (Toulouse), he has joined the Institut Langevin (CNRS \& ESPCI, Paris) in 2017 as postdoctoral researcher co-funded by Greenerwave and the French Ministère des Armées, Direction Générale de l’Armement. Jean-Baptiste’s current activities focus on the theory and applications of spatial microwave modulation with Greenerwave technology and Wavefront shaping in electromagnetic reverberating media.
\end{IEEEbiographynophoto}

\begin{IEEEbiographynophoto}{Vladislav Popov} was born in Minsk, Belarus, on February 13, 1994. In 2017 he received his Master Degree (with distinction) from the Department of Theoretical Physics and Astrophysics, Belarusian State University, Belarus.
	In 2020 he received a Ph.D. degree in Electrical Engineering at CentraleSup\'elec (University Paris-Saclay, France) on developing electrically reconfigurable metasurface-based conformal antennas.
	His current research interest include advanced concepts in metamaterials and metasurfaces, design and homogenization techniques of metamaterials, reconfigurable and conformal sparse metasurfaces. 
	He serves as a reviewer for various journals including IEEE Transactions on Antennas and Propagation, Physical Review Letters, and Advanced Optical Materials.
	
	He received two awards of The special fund of the President of the Republic of Belarus for the social support of gifted students  (in 2011 and 2014).
	In 2016 he received the $3^\textup{rd}$ prize for Student Paper Competition at the Tenth International Congress on Advanced Electromagnetic Materials in Microwaves and Optics: ``Metamaterials2016''.
	
\end{IEEEbiographynophoto}

\begin{IEEEbiographynophoto}{Mikhail A. Odit} born in St.Petersburg, Russia in 1982. He received diploma of microwave engineer in 2005 from the Electrotechnical University LETI, St. Petersburg, Russia. He obtained his Ph.D. degree in 2010 working in the field on dielectric bulk metamaterials. He hold a position of the associate professor in the Microwave Electronics Department of the University LETI. Since 2015 Mikhail worked as a research fellow in the metamaterials laboratory of ITMO University (St. Petersburg, Russia) conducting research in the field of microwave metamaterials and metasurfaces. 
His current research interest lies in the field of mmWave metasurfaces for antenna applications. 
Mikhail is a co-author of over 14 paper in peer-reviewed journals and co-author of 2 chapters in books. 

In 2019 he joined Greenerwave as a senior research engineer. He manages the projects related to application of metasurfaces for satellite communication antennas and 5G.
\end{IEEEbiographynophoto}	

\begin{IEEEbiographynophoto}
{Vladimir Lenets} received the Master’s degree from the Department of Physics, ITMO University, St. Petersburg, Russia under the supervision of Prof. Stanislav Glybovski working on focusing circular-polarization beam splitter based on a self-complementary metasurface. Also, during his Master’s degree, he was involved in other projects related to metasurfaces for antenna application and experimental investigation of metasurfaces properties. He joined Institut Langevin in 2020 as an engineer and is mainly engaged in wavefront shaping in complex media.
\end{IEEEbiographynophoto}	

\begin{IEEEbiographynophoto}	
{Geoffroy Lerosey}  is the co-founder and the inventor of the concepts behind Greenerwave with Mathias Fink. He is on leave from academia to fully support the company’s development and bring his scientific expertise.
Geoffroy earned an engineering degree from ESPCI Paris, a Master’s degree in electronics from Université Pierre et Marie Curie and a PhD in Physics from Université Paris Diderot. He joined University of California at Berkeley for Postdoctoral researches working mainly on metamaterials and plasmonics. Coming back to France, Geoffroy was appointed by French main academic research organization CNRS in 2008 and started a group at Institut Langevin (CNRS \& ESPCI Paris). Geoffroy’s researches are in metamaterials and metasurfaces, time reversal and signal processing, subwavelength imaging and focusing techniques, wavefront shaping in optics and RF, photonic and phononic crystals, reverberating and locally resonant media, and span all domains of wave physics from acoustics to optics.
Geoffroy has been invited more than 80 times at international conferences, and has given invited seminars in many universities worldwide. He supervised 8 PhD students, 6 postdocs and 20 Master students. His research led 100 scientific articles, 15 patents and 2 startups.
\end{IEEEbiographynophoto}

\end{document}